\documentstyle[12pt,psfig]{article}

\def\simlt{\stackrel{<}{{}_\sim}}
\def\simgt{\stackrel{>}{{}_\sim}}
\def\be{\begin{equation}}
\def\ee{\end{equation}}
\def\bear{\be\begin{array}}
\def\eear{\end{array}\ee}
\def\bea{\begin{eqnarray}}
\def\eea{\end{eqnarray}}
\catcode`@=11
\newtoks\@stequation
\def\subequations{\refstepcounter{equation}%
  \edef\@savedequation{\the\c@equation}%
  \@stequation=\expandafter{\theequation}
  \edef\@savedtheequation{\the\@stequation}
  \edef\oldtheequation{\theequation}%
  \setcounter{equation}{0}%
  \def\theequation{\oldtheequation\alph{equation}}}
\def\endsubequations{\setcounter{equation}{\@savedequation}%
  \@stequation=\expandafter{\@savedtheequation}%
  \edef\theequation{\the\@stequation}\global\@ignoretrue

\noindent}
\catcode`@=12
\catcode`\@=11
\long\def\@makefntext#1{
\protect\noindent \hbox to 3.2pt {\hskip-.9pt
$^{{\ninerm\@thefnmark}}$\hfil}#1\hfill}		

 \def\@makefnmark{\hbox to 0pt{$^{\@thefnmark}$\hss}}  

\def\ps@myheadings{\let\@mkboth\@gobbletwo
\def\@oddhead{\hbox{}
\rightmark\hfil\ninerm\thepage}
\def\@oddfoot{}\def\@evenhead{\ninerm\thepage\hfil
\leftmark\hbox{}}\def\@evenfoot{}
\def\sectionmark##1{}\def\subsectionmark##1{}}

\newcounter{sectionc}\newcounter{subsectionc}\newcounter{subsubsectionc}
\renewcommand{\section}[1] {\vspace{0.6cm}\addtocounter{sectionc}{1}
\setcounter{subsectionc}{0}\setcounter{subsubsectionc}{0}\noindent
	{\bf\thesectionc. #1}\par\vspace{0.4cm}}
\renewcommand{\subsection}[1] {\vspace{0.6cm}\addtocounter{subsectionc}{1}
	\setcounter{subsubsectionc}{0}\noindent
	{\it\thesectionc.\thesubsectionc. #1}\par\vspace{0.4cm}}
\renewcommand{\subsubsection}[1]
{\vspace{0.6cm}\addtocounter{subsubsectionc}{1}
	\noindent {\rm\thesectionc.\thesubsectionc.\thesubsubsectionc.
	#1}\par\vspace{0.4cm}}

\newcounter{appendixc}
\newcounter{subappendixc}[appendixc]
\newcounter{subsubappendixc}[subappendixc]

\renewcommand{\appendix}[1] {\vspace{0.6cm}
        \refstepcounter{appendixc}
        \setcounter{figure}{0}
        \setcounter{table}{0}
        \setcounter{equation}{0}
        \renewcommand{\thefigure}{\Alph{appendixc}.\arabic{figure}}
        \renewcommand{\thetable}{\Alph{appendixc}.\arabic{table}}
        \renewcommand{\theappendixc}{\Alph{appendixc}}
        \renewcommand{\theequation}{\Alph{appendixc}.\arabic{equation}}
        \noindent{\bf Appendix \theappendixc #1}\par\vspace{0.4cm}}

\def\abstracts#1{{
	\centering{\begin{minipage}{30pc}\tenrm\baselineskip=12pt\noindent
	\centerline{\tenrm ABSTRACT}\vspace{0.3cm}
	\parindent=0pt #1
	\end{minipage}}\par}}

\renewenvironment{thebibliography}[1]
	{\begin{list}{\arabic{enumi}.}
	{\usecounter{enumi}\setlength{\parsep}{0pt}
\setlength{\leftmargin 1.25cm}{\rightmargin 0pt}
	 \setlength{\itemsep}{0pt} \settowidth
	{\labelwidth}{#1.}\sloppy}}{\end{list}}

\newcounter{itemlistc}
\newcounter{romanlistc}
\newcounter{alphlistc}
\newcounter{arabiclistc}

\newcommand{\fcaption}[1]{
        \refstepcounter{figure}
        \setbox\@tempboxa = \hbox{\tenrm Fig.~\thefigure. #1}
        \ifdim \wd\@tempboxa > 6in
           {\begin{center}
        \parbox{6in}{\tenrm\baselineskip=12pt Fig.~\thefigure. #1}
            \end{center}}
        \else
             {\begin{center}
             {\tenrm Fig.~\thefigure. #1}
              \end{center}}
        \fi}

\newcommand{\tcaption}[1]{
        \refstepcounter{table}
        \setbox\@tempboxa = \hbox{\tenrm Table~\thetable. #1}
        \ifdim \wd\@tempboxa > 6in
           {\begin{center}
        \parbox{6in}{\tenrm\baselineskip=12pt Table~\thetable. #1}
            \end{center}}
        \else
             {\begin{center}
             {\tenrm Table~\thetable. #1}
              \end{center}}
        \fi}

\def\@citex[#1]#2{\if@filesw\immediate\write\@auxout
	{\string\citation{#2}}\fi
\def\@citea{}\@cite{\@for\@citeb:=#2\do
	{\@citea\def\@citea{,}\@ifundefined
	{b@\@citeb}{{\bf ?}\@warning
	{Citation `\@citeb' on page \thepage \space undefined}}
	{\csname b@\@citeb\endcsname}}}{#1}}

\newif\if@cghi
\def\cite{\@cghitrue\@ifnextchar [{\@tempswatrue
	\@citex}{\@tempswafalse\@citex[]}}
\def\citelow{\@cghifalse\@ifnextchar [{\@tempswatrue
	\@citex}{\@tempswafalse\@citex[]}}
\def\@cite#1#2{{$\null^{#1}$\if@tempswa\typeout
	{IJCGA warning: optional citation argument
	ignored: `#2'} \fi}}

\def\fnt#1#2{\footnotetext{\kern-.3em
	{$^{\mbox{\sevenrm #1}}$}{#2}}}

 1
 1
 1

\font\tenbf=cmbx10
\font\tenrm=cmr10
\font\tenit=cmti10

\font\ninerm=cmr9

\begin{document}


\begin{titlepage}

\title{\large {\bf Bounds on the Higgs mass in the Standard Model and
\phantom{Bounds on the Higgs mass in the Standard Model and}
Minimal Supersymmetric Standard Model} \thanks{Based on talks given
at {\it Physics from Planck scale to electroweak scale}, Warsaw,
Poland, 21-24 September 1994, and {\it Padua meeting of the European
network on: "Phenomenology of the Standard Model and alternatives for
present and future high energy colliders"}, Padua, Italy,
4-5 November 1994.}}

\author{ \\
{\large M. Quir\'os}\\
\\
{\large CERN, TH Division, CH-1211 Geneva 23, Switzerland}}
\date{}
\maketitle
\def\baselinestretch{1.15}
\vspace{1cm}
\begin{abstract}
\noindent
We present bounds on the Higgs mass in the Standard Model
and in the Minimal Supersymmetric Standard Model using the effective
potential with next-to-leading logarithms resummed by
the renormalization group equations, and physical (pole) masses for
the top quark and Higgs boson. In the Standard Model we obtain lower
bounds from stability requirements: they depend on the top mass
and the cutoff scale. In the Minimal Supersymmetric Standard
Model we obtain upper bounds which depend on the top mass and the
scale of supersymmetry breaking.
A Higgs mass measurement could discriminate,
depending on the top mass, between the two models. Higgs discovery at
LEP-200 can put an upper bound on the scale of new physics.
\end{abstract}

\thispagestyle{empty}

\vspace{3.cm}
\leftline{}
\leftline{CERN--TH.7507/94}
\leftline{November 1994}
\leftline{}

\vskip-22.cm
\rightline{}
\rightline{CERN--TH.7507/94}
\rightline{IEM--FT--95/94}
\rightline{hep-ph/9411403}
\vskip3in

\end{titlepage}

\newpage


\centerline{\tenbf BOUNDS ON THE HIGGS MASS IN THE STANDARD MODEL AND}
\baselineskip=16pt
\centerline{\tenbf MINIMAL SUPERSYMMETRIC STANDARD MODEL}
\vspace{0.8cm}
\centerline{\tenrm MARIANO QUIROS \footnote{Work partly supported
by CICYT under contract
AEN94-0928, and by the European Union under contract No. CHRX-CT92-0004.}}
\baselineskip=13pt
\centerline{\tenit TH Division, CERN, CH-1211}
\baselineskip=12pt
\centerline{\tenit Geneva 23, Switzerland}
\vspace{1.5cm}
\abstracts{We present bounds on the Higgs mass in the Standard Model
and in the Minimal Supersymmetric Standard Model using the effective
potential with next-to-leading logarithms resummed by
the renormalization group equations, and physical (pole) masses for
the top quark and Higgs boson. In the Standard Model we obtain lower
bounds from stability requirements: they depend on the top mass
and the cutoff scale. In the Minimal Supersymmetric Standard
Model we obtain upper bounds which depend on the top mass and the
scale of supersymmetry breaking.
A Higgs mass measurement could discriminate,
depending on the top mass, between the two models. Higgs discovery at
LEP-200 can put an upper bound on the scale of new physics.}

\vfil
\rm\baselineskip=14pt

\section{Introduction}

In view of Higgs searches at future colliders (in particular at LEP-200)
it is extremely important to compute theoretical bounds on the Higgs
mass as accurately as possible. Indeed, the very measurement of the
Higgs mass, apart from confirming our knowledge of renormalizable field
theories, can shed some light on the
particular model that Nature has chosen at the TeV scale.

The two most
appealing models are: the Standard Model (SM), which is being confirmed
at LEP with $\sim$ 1\% precision, and its minimal supersymmetric extension,
the Minimal Supersymmetric Standard Model (MSSM), which is very well
motivated theoretically (it helps in technically solving the gauge
hierarchy problem) but has received no experimental support
from present colliders (TEVATRON, LEP). On the contrary
what present colliders
are doing is to establish lower bounds on all supersymmetric parameters,
and therefore sending the scale of supersymmetry breaking to higher values.

The direct confirmation or exclusion of the MSSM will probably have to wait
for the LHC. In the meantime, while present and future accelerators
will continue to
rise the bounds on supersymmetric parameters, we would like to know whether
the discovery of the Higgs boson and the measurement of its mass could,
apart from confirming the SM, give information about the possibility that
the SM is embedded in an extended electroweak theory: in particular the
MSSM. On the one hand, if all supersymmetric parameters are at the TeV scale,
the lightest Higgs boson of the MSSM should have the same couplings to
ordinary matter than the SM Higgs. On the other hand LEP-200 will cover a
range of Higgs masses up to $\sim M_Z$ and direct detection of supersymmetry
will not happen if supersymmetric masses are not $\simlt M_Z$. If this is
the case, the best tool LEP-200 will have to uncover new physics is through
the Higgs search and the measurement of its mass.

In this talk we will review the theoretical knowledge we have on the SM
Higgs mass and on the MSSM lightest Higgs boson mass. In Section 2 we
will present lower bounds on the SM Higgs mass from the requirement of
stability of the effective potential, i.e. from the requirement
that we do not live in a metastable minimum. We will see that it imposes
strong lower bounds on the mass of the Higgs boson, $M_H$, which will
be a function of the top mass $M_t$ and of the SM cutoff $\Lambda$.
In Section 3 we will review the upper bounds on the lightest Higgs boson
mass $M_H$ that will depend on various supersymmetric parameters and
$M_t$. In Section 4 we will draw some conclusions concerning the possible
detection of the Higgs at LEP-200.

\section{Lower bounds in the Standard Model: stability bounds}

The vacuum stability requirement in the SM
imposes a severe lower bound on the mass of the Higgs boson
$M_H$~$^{1-5}$,
which depends on the mass of the top quark $M_t$ and on the cut-off
$\Lambda$ beyond which the SM is no longer valid. Roughly speaking, this is
due to the fact that the top Yukawa coupling $h_t$ drives the quartic
coupling of the Higgs potential $\lambda$ from its initial value at
$M_Z$ (which determines the Higgs mass) to negative values at
large scales, thus destabilizing the standard electroweak vacuum.

In previous works, the stability
bound was obtained from the tree level potential,
improved by one-loop or two-loop renormalization group equations (RGE) for
the $\beta$- and $\gamma$-functions of the running couplings, masses and the
$\phi$-field~\cite{LSZ,Sher}. However, it has been shown
that the one-loop corrections to the Higgs potential are important in order
to fix the boundary conditions for the electroweak breaking and calculate
the Higgs mass in a consistent and scale-independent way. As we will see,
they are also significant to properly understand the whole structure of the
potential. Typically we find that the lower bound on $M_H$ is ${\cal
O}$(10 GeV) lower than in previous estimates~\cite{Sher}.

The renormalization group improved effective potential of the SM, $V$, can
be written in the 't Hooft-Landau gauge and the $\overline{MS}$ scheme
as~\cite{Einhorn}
\be
\label{veff}
V[\mu(t),\lambda_i(t);\phi(t)]\equiv V_0 + V_1 + \cdots \;\; ,
\ee
where $\lambda_i\equiv (g,g',\lambda,h_t,m^2)$ runs over all
dimensionless and dimensionful couplings, and $V_0$, $V_1$ are respectively
the tree level potential and the one-loop correction, namely

\subequations{
\be
\label{v0}
V_0=-{\displaystyle\frac{1}{2}}m^2(t)\phi^2(t) +
{\displaystyle\frac{1}{8}}\lambda(t)\phi^4(t),
\ee
\be
\label{v1}
V_1={\displaystyle\sum_{i=1}^5}{\displaystyle\frac{n_i}{64\pi^2}}
M_i^4(\phi)\left[\log{\displaystyle\frac{M_i^2(\phi)}{\mu^2(t)}}
-c_i\right]+\Omega(t),
\ee}
\endsubequations
where $M_i^2(\phi,t)=\kappa_i\phi^2(t)-\kappa_i'$
(1~$\equiv$~$W$, 2~$\equiv$~$Z$, 3~$\equiv$~top,
4~$\equiv$~Higgs and 5~$\equiv$~Goldstones)
are the tree-level expressions for the masses of the
particles that enter in the one-loop radiative corrections, and
$n_1=6$, $\kappa_1=\frac{1}{4}g^2(t)$, $\kappa_1'=0$,
$c_1=\frac{5}{6}$;
$n_2=3$, $\kappa_2=\frac{1}{4}[g^2(t)+g'^2(t)]$,
$\kappa_2'=0$, $c_2=\frac{5}{6}$;
$n_3=-12$, $\kappa_3=\frac{1}{2}h_t^2(t)$, $\kappa_3'=0$,
$c_3=\frac{3}{2}$;
$n_4=1$, $\kappa_4=\frac{3}{2}\lambda(t)$,
$\kappa_4'=m^2(t)$,
$c_4=\frac{3}{2}$;
$n_5=3$, $\kappa_5=\frac{1}{2}\lambda(t)$,
$\kappa_5'=m^2(t)$, and
$c_5=\frac{3}{2}$.
$\Omega(t)\equiv \Omega[\lambda_i(t),\mu(t)]$ is
the one-loop contribution to the cosmological constant~\cite{Einhorn},
which will turn out to be irrelevant in our calculation.

In the previous expressions the parameters $\lambda(t)$ and $m(t)$ are the SM
quartic coupling and mass, whereas $g(t)$, $g'(t)$, $h_t(t)$ are the SU(2),
U(1) and top Yukawa couplings respectively. All of them are running with
the RGE. The running of the Higgs field is
$\phi(t)=\xi(t) \phi_c$,
$\phi_c$ being the classical field and
$\xi(t)=\exp\{-\int^t_0 \gamma(t')dt'\}$,
where $\gamma(t)$ is the Higgs field anomalous dimension.
Finally the scale $\mu(t)$ is related to the
running parameter $t$ by
$\mu(t)=\mu e^t$,
where $\mu$ is a fixed scale, we will take equal to the physical
$Z$ mass, $M_Z$.

It has been shown~\cite{Lpot} that the $L$-loop  effective potential
improved by ($L$+1)-loop RGE resums all
$L$th-to-leading logarithm contributions.
Consequently, we will consider all the $\beta$- and $\gamma$-functions of
the previous parameters to two-loop order, so that our calculation will
be valid up to next-to-leading logarithm approximation.

As has been pointed out~\cite{Einhorn}, working with $\partial V/
\partial \phi$ (and higher derivatives) rather than with $V$ itself allows
us to ignore the cosmological constant~\footnote{This holds even
if we choose $\mu(t)$ to be a function of the $\phi$-field since the
scale-invariant properties of $V$ allow
the substitution to be performed either
before or after taking the derivative~\cite{CEQR}
$\partial/\partial\phi$.} \phantom{a}term $\Omega$. In fact,
the structure of the potential
can be well established once we have determined the values of $\phi$, say
$\phi_{{\mathrm ext}}$, in which $V$ has extremals (maxima or minima).
Thus we only need to evaluate $\partial V/
\partial \phi$ and $\partial^2 V/
\partial \phi^2$.

The structure of maxima and minima of $V$ for large $\phi$ can be
evaluated at a scale
$\mu(t)$ within the region where $V$ is scale-invariant. As was
previously discussed~\cite{CEQR},
$\mu(t)=\phi(t)$ is always a correct
choice (other choices, such that $\mu(t)=\phi(t)/2$, are equally valid and lead
to essentially the same results). Then the extremal condition,
neglecting the Higgs and Goldstone contributions, reads~\cite{CEQ}
\be
\label{phiext}
\phi^2_{{\mathrm ext}}=\frac{2m^2}{\tilde{\lambda}},
\ee
\bea
\label{lambdahat}
\tilde{\lambda}&=&\lambda-
{\displaystyle\frac{1}{16\pi^2}}\left\{6h_t^4\left[\log
{\displaystyle\frac{h_t^2}{2}}-1\right]-
{\displaystyle\frac{3}{4}}g^4\left[\log
{\displaystyle\frac{g^2}{4}}-{\displaystyle\frac{1}{3}}
\right]\right.\nonumber \\
&&\nonumber \\
&-&\left.{\displaystyle\frac{3}{8}}\left(g^2+g'^2\right)^2
\left[\log{\displaystyle\frac{
\left(g^2+g'^2\right)}{4}}-{\displaystyle\frac{1}{3}}\right]
\right\},
\eea
[all quantities in (\ref{phiext},\ref{lambdahat}) are evaluated at
$\mu(t)=\phi_{{\mathrm ext}}(t)$]. From (\ref{phiext}) we see that, if $V$
develops an extremal for large values of $\phi$, this must occur for
a value of $\phi$ such that
\be
\label{lext}
0<\tilde{\lambda}[\mu(t)=\phi(t)]\ll 1.
\ee
On the other hand, for large values of $\phi$
the second derivative of the potential (\ref{veff})
can be very accurately expressed as~\cite{CEQ}
\bea
\label{secblarge}
\left.\frac{\partial^2 V}{\partial \phi^2(t)}\right|_{\phi(t)=
\phi_{{\mathrm ext}}(t)}
&=&\frac{1}{2}(\beta_{\lambda}-4\gamma\lambda)\phi^2(t),
\eea
where $\beta_{\lambda}$ is the one-loop
$\beta$-function.
Since near the extremum $\lambda$ is very
small, we see from (\ref{secblarge}) that depending on the sign of
$\beta_{\lambda}$ we will have a maximum or a minimum.

\begin{figure}[htb]
\centerline{
\psfig{figure=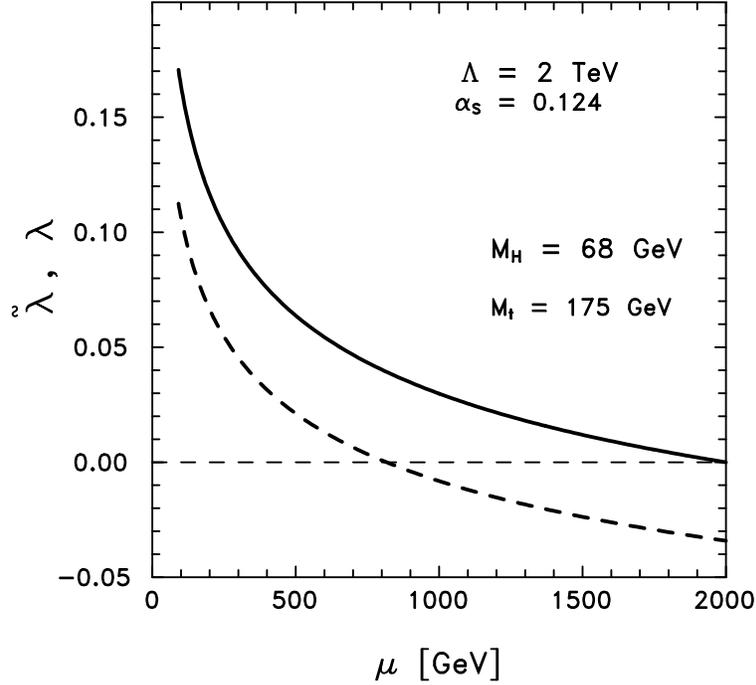,height=10cm,bbllx=5cm,bblly=5cm,bburx=18cm,bbury=17cm}}
\caption[0]{Plot of $\lambda$ (dashed line) and
$\tilde{\lambda}$ (solid line) as a function of the scale
$\mu(t)$ for $M_t=175$ GeV, $\Lambda=2$ TeV, $M_H=68$ GeV and
$\alpha_S=0.124$.}
\label{lambdas}
\end{figure}

\begin{figure}[htb]
\centerline{
\psfig{figure=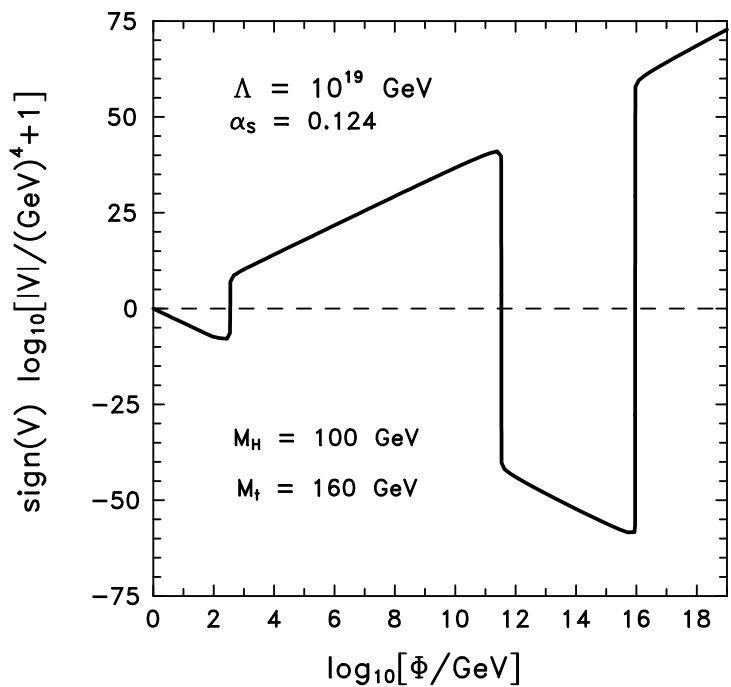,height=10cm,bbllx=5cm,bblly=5cm,bburx=18cm,bbury=17cm}}
\caption[0]{Plot of the effective potential $V(\phi)$ corresponding
to $M_t=160$ GeV, $M_H=100$ GeV, $\alpha_S=0.124$ and
$\Lambda=10^{19}$ GeV, represented in a convenient choice of units.}
\label{potential}
\end{figure}

We have illustrated these features in Fig.~\ref{lambdas}
with a typical example.
It represents the evolution of $\lambda$ (dashed line) and
$\tilde{\lambda}$ (solid line)
with $\mu(t)$. It is worth noticing that they do not cross the
horizontal axis at the same value of $\mu(t)$, but they differ by a
relatively large amount. This is important since the point
where the maximum of the potential is located, say $\phi_{MAX}$, does
correspond~\cite{CEQ}
to $\tilde{\lambda}\sim 0$ rather than ${\lambda}\sim 0$.
This is so in Fig.~\ref{potential}, where the scalar
potential, $V(\phi)$, has been represented for a typical choice
of parameters.
Notice that for values of $\phi$ very
slightly higher than $\phi_{MAX}$, the potential is negative and much
deeper than the electroweak minimum. This is simply because for values
of $\mu(t)$ just beyond $\mu_{MAX}=\phi_{MAX}$, the value
of $\tilde{\lambda}$ becomes negative and the
potential is dominated by the contribution
$\frac{1}{8}\tilde{\lambda}\phi^4$.
Consequently, a sensible criterion for a model to
be safe is to require one of the two following conditions:
{\bf\it a)}~The potential has no maximum; {\bf\it b)}~The
maximum occurs for $\phi_M>\Lambda$.
In the following we will assume $\Lambda\leq
10^{19}$ GeV. With this criterion [in particular condition ({\it b})]
we see that the model represented in Fig.~\ref{potential} is acceptable for
$\Lambda\leq 2.7\times 10^{11}$ GeV.
Beyond this scale, the stability of the vacuum requires the appearance
of new physics. Note from this discussion that conditions
({\it a}), ({\it b}) are not equivalent to require
$\lambda(\mu)>0$ for $\mu(t)<\Lambda$, as is usually done.
Instead, the significant parameter is $\tilde{\lambda}$ rather than
${\lambda}$.

The running Higgs mass, $m_H^2(t)$, defined as the curvature of the
scalar potential at the minimum,
can be readily obtained from
\be
m_H^2(t^*)=\left.\frac{\partial^2V}{\partial\phi^2(t^*)}\right|_{\phi(t^*)
=\langle\phi(t^*)\rangle},
\ee
where $t^*$ is the scale at which we define the electroweak
minimum~\cite{CEQR}.
The scale invariance of the second derivative of the potential,
$\frac{\partial}{\partial t}\left[\xi^2(t)\frac{\partial^2V}
{\partial\phi^2(t)}
\right]=0$,
allows us to write $m_H^2(t)$ at any arbitrary scale
\be
\label{mhrun}
m_H^2(t)=m_H^2(t^*){\displaystyle\frac{\xi^2(t^*)}{\xi^2(t)}}.
\ee
The physical (pole) Higgs mass, $M_H^2$, is then given by
\be
\label{MHphys}
M_H^2= m_{H}^2(t) +{{\mathrm Re}}[\Pi (p^2=M_H^2)-\Pi (p^2=0)],
\ee
where $\Pi(p^2)$ is the renormalized self-energy of the
Higgs boson~\cite{CEQR}
(the $t$-dependence drops out from (\ref{MHphys})).

\begin{figure}[htb]
\centerline{
\psfig{figure=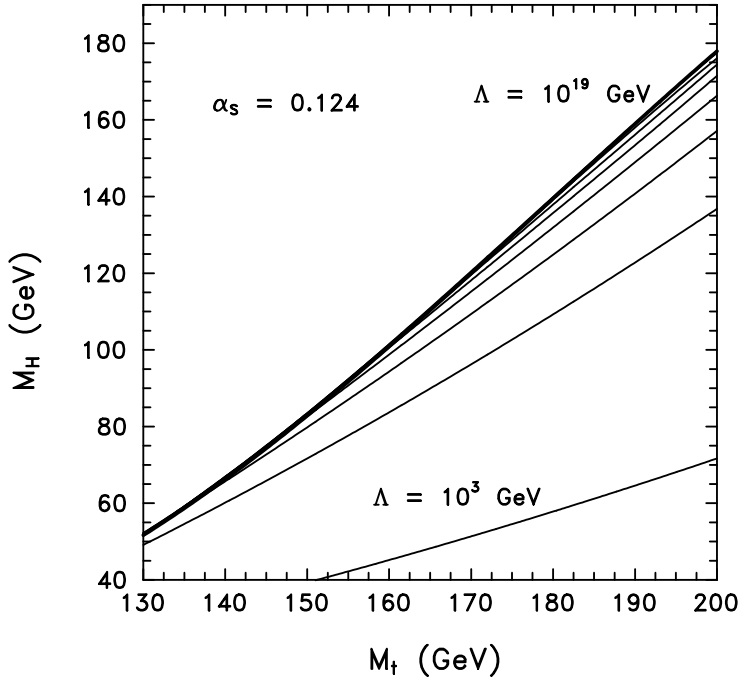,height=10cm,bbllx=5cm,bblly=5cm,bburx=18cm,bbury=17cm}}
\caption[0]{SM lower bound on $M_H$ as a function of $M_t$ for
$\alpha_S(M_Z)=0.124$ and different values of $\Lambda$ in the range
$10^3$ GeV $\leq \Lambda \leq 10^{19}$ GeV. The values of $\Lambda$ for
consecutive curves differ by two orders of magnitude.}
\label{cotas}
\end{figure}

As has been stated above, the choice of $\mu^*$, i.e. the scale at
which we evaluate the minimum conditions, is not important for physical
quantities, provided it is within a (quite wide) region around the optimal
value~\cite{CEQ}.
The lack of flatness of $M_H$ reflects the effect of
all non-considered (higher-order) contributions in the calculation
and, therefore, it is a measure of the total error in our estimate
of $M_H$. We deduced~\cite{CEQ} that the error is typically
$\simlt$ 3 GeV, which is the uncertainty we should assign to our
results.
Had we performed the
previous calculations just with the (RGE-improved) tree-level part
of $V$ in Eq.~(\ref{veff})~\cite{LSZ,Sher},
the Higgs mass would have a strong dependence on $\mu^*$.
Choosing~\cite{LSZ,Sher} $\mu^*=M_Z$
results in an error in the estimate of $M_H$,
whose precise value depends on the top
mass and is typically of ${\cal O}(10\ {{\mathrm GeV}})$,
showing the need of a more careful treatment of the
problem, as the one exposed above~\cite{CEQ}.

Finally, let us note that in the previous equations the top Yukawa
coupling $h_t(t)$ enters in several places. Therefore, the Higgs
mass depends on the boundary condition chosen for $h_t(t)$, and thus
on the top mass $M_t$. However the running top mass, defined as
$m_t(t)=vh_t(t)$, does not coincide with the physical (pole) mass
$M_t$. In the Landau gauge the relationship between the running $m_t$ and the
physical (pole) mass $M_t$ is given by~\cite{gray}
\be
\label{mtphys}
M_t=\left\{1+{\displaystyle\frac{4}{3}}
{\displaystyle\frac{\alpha_S(M_t)}{\pi}}+
\left[16.11-1.04\sum_{i=1}^{5}\left(1-\frac{M_i}{M_t}\right)\right]
\left(\frac{\alpha_S(M_t)}{\pi}\right)^2
\right\} m_t(M_t),
\ee
where $M_i$, $i=1,\ldots,5$ represent the masses of the five lighter quarks.

As has become clear from the  previous discussion,
the lower bound on $M_H$ is a function of $M_t$ and
$\Lambda$. However,
apart from the previously estimated error $\simlt$ 3 GeV
in our calculation, there is an additional source of uncertainty
coming from the value of $\alpha_s$, which enters in several places
in the previous calculation. The most recent estimate of $\alpha_s$
gives
\be
\label{alphas}
\alpha_s=0.124\pm 0.006.
\ee
Using the central value of (\ref{alphas}), we have represented in
Fig.~\ref{cotas} the lower bound on $M_H$ as a function of $M_t$ for different
values of $\Lambda$. The form of the curves is easily understandable
from the previous discussion. In Fig.~\ref{comparacion},
we have fixed $\Lambda$ at its maximum value, $\Lambda=10^{19}$ GeV, and
represented the lower bound on $M_H$ for the central value of
$\alpha_s$ in (\ref{alphas}) (diagonal solid line) and the two
extreme values (diagonal dashed lines).

If we use the recent evidence for the top quark
production at CDF with a mass $M_t=174\pm 17$ GeV~\cite{cdf},
we obtain the following lower bound on $M_H$:
\be
\label{174bound}
M_H> 128\pm 33\ {{\mathrm GeV}},
\ee
i.e. $M_H>$ 95 GeV (1$\sigma$).
If the Higgs is observed in the present or forthcoming accelerators
with a mass below the bound of Eq.~(\ref{174bound}), this would be a
clear signal of new physics beyond the SM.

Comparing these bounds with previous evaluations~\cite{Sher},
we see that our values of $M_H$ are
lower by an amount increasing with $M_t$,
going from $\sim 5$ GeV for $M_t\sim$~$130$~GeV to $\sim 15$ GeV for
$M_t \sim 200$ GeV.
As has been discussed previously
the main reason of this difference is the way in which the Higgs mass
was previously computed~\cite{Sher}. Accordingly, our results give more
room to the Higgs mass in the framework of the Standard Model.

\section{Upper bounds in the Minimal Supersymmetric Standard Model}

%
The MSSM has an extended Higgs sector with two Higgs doublets with
opposite hypercharges: $H_1$, responsible for the mass of the charged
leptons and the down-type quarks, and $H_2$, which gives a mass to the
up-type quarks. After the Higgs mechanism there remain three physical scalars,
two CP-even and one CP-odd Higgs bosons. In particular, the lightest
CP-even Higgs boson mass satisfies the tree-level bound
\begin{equation}
m_H^2\leq M_Z^2\cos^2 2\beta,
\label{bound}
\end{equation}
where $\tan\beta=v_2/v_1$ is the ratio of the Vacuum Expectation Values
(VEVs) of the neutral components of the two Higgs
fields $H_2$ and $H_1$. Relation (\ref{bound})
implies that $m_H^2<M_Z^2$, for any value of $\tan\beta$, which, in turn,
implies that it should be found at LEP-200~\cite{lep200}. However, the
tree-level relation (\ref{bound}) is spoiled by one-loop radiative
corrections, which were computed by several  groups using:
the effective potential approach~\cite{eff}, diagrammatic methods~\cite{diag}
and renormalization group (RG) techniques~\cite{rg}. All
methods found excellent agreement with one another and large radiative
corrections, mainly controlled by the top Yukawa coupling, which
could make the lightest CP-even Higgs boson escape experimental
detection at LEP-200. In particular, the RG approach (which will be
followed in this talk) is based on the
fact that supersymmetry decouples and, below the scale of
supersymmetry breaking $M_S$, the effective theory is the SM,
with some matching conditions at $M_S$.
Assuming $M_Z^2\ll M_S^2$ the tree-level bound
(\ref{bound}) is saturated at the scale $M_S$ and the effective SM at
scales between $M_Z$ and $M_S$ contains the Higgs doublet
$H=H_1\cos\beta +i \sigma_2 H_2^*\sin\beta$,
with a quartic coupling taking, at the scale $M_S$,
the (tree-level) value of
\be
\label{lms}
\lambda={\displaystyle\frac{1}{4}} (g^2+g'^2)\cos^2 2\beta.
\ee
In these analyses~\cite{rg} the Higgs mass was considered at
the tree level, improved by one-loop renormalization group equations
(RGE) in the $\gamma $- and
$\beta $-functions, thus collecting all leading logarithm corrections.

Since the relative size of one-loop corrections to the Higgs mass is
large (mainly for large top quark mass and/or small tree-level Higgs
mass) it was compelling to analyse them at the two-loop level. A
first step in that direction was given some time ago~\cite{eq} where
two-loop RGE-improved tree-level Higgs masses were considered. It was
found that two-loop corrections were negative and small.
The Higgs mass received all leading logarithm and part of the
next-to-leading logarithm corrections~\cite{eq}.
As we have described in Section 2, for
fully taking into account all next-to-leading logarithm corrections the
one-loop effective potential (improved by two-loop RGE) is needed.

The tree-level quartic
coupling (\ref{lms}) receives one-loop threshold contributions at the
$M_S$ scale. These are given by
\be
\label{deltal}
\Delta\lambda={\displaystyle\frac{3 h_t^4}{16
\pi^2}}{\displaystyle\frac{X_t^2}{M_S^2}}\left(2-
{\displaystyle\frac{X_t^2}{6 M_S^2}}\right),
\ee
where $h_t$ is the top Yukawa coupling in the SM and
$X_t=A_t+\mu\cot\beta$
is the stop mixing.

The correction (\ref{deltal}) has a maximum for $X_t^2=6M_S^2$.
For that reason, in our numerical applications we will take
$X_t^2=6M_S^2$,
i.e. maximal threshold effect. Notice also that $X_t^2=6M_S^2$
is barely consistent with the bound from colour-conserving
minimum~\cite{color}, so that
the case of maximal threshold really represents
a particularly extreme situation.
In addition to the previous effect, there appear effective
higher-order operators ($D\geq 6$), which for $M_S\geq 1$ TeV
turn out to be negligible~\cite{CEQR}.

\begin{figure}[htb]
\centerline{
\psfig{figure=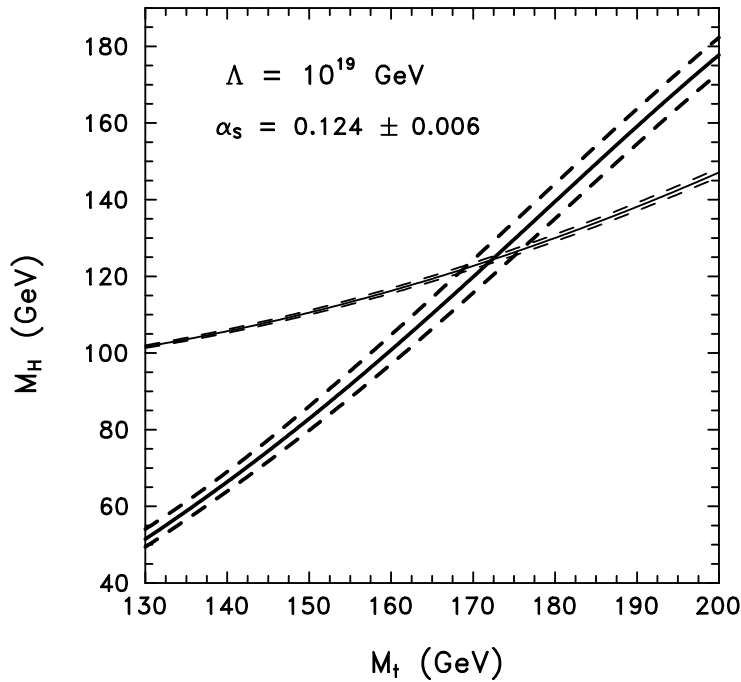,height=10cm,bbllx=5cm,bblly=5cm,bburx=18cm,bbury=17cm}}
\caption[0]{Diagonal (thick) lines: SM lower bound on $M_H$ as a
function of $M_t$ for $\Lambda=10^{19}$ GeV and $\alpha_s=0.124$
(solid line), $\alpha_s=0.118$ (upper dashed line), $\alpha_s=0.130$
(lower dashed line). Transverse (thin) lines: MSSM upper bounds on
$M_H$ for $\Lambda_S=1$ TeV and $\alpha_s$ as in the diagonal lines.}
\label{comparacion}
\end{figure}

Upper bounds on the lightest Higss boson mass in the MSSM
depend therefore on three supersymmetric parameters
(besides $M_t$): $M_{S}$ (from naturality reasons
$M_{S}\simlt 1$~TeV~\cite{fine}), $\tan\beta$ and
$X_t=A_t+\mu/\tan\beta$, which is responsible for the
threshold correction to the Higgs quartic coupling.
The larger the threshold correction and $\tan\beta$, the less stringent
the supersymmetric bounds. Therefore, the most conservative situation takes
place considering maximum threshold correction (which is achieved for
$X_t^2=6M_{S}^2$) and $\tan\beta=\infty$. Likewise, the larger
$M_{S}$, the less stringent the bounds; but, as mentioned
above, it is not sensible to consider $M_{S}$ much larger
than $1$ TeV. Consequently, to be on the safe side, we have
represented in Fig.~\ref{comparacion}
the MSSM upper bounds (transverse solid and
dashed lines), as recently obtained up
to next-to-leading-log order~\cite{CEQR}, in the most
conservative situation with $M_{S}=1$ TeV.

Two papers have recently tried to
incorporate radiative corrections to the Higgs mass up to the
next-to-leading order, and with qualitatively different results.
Using the RG approach,
positive and large next-to-leading corrections, with
respect to the one-loop results, were found
by Kodaira, Yasui and Sasaki (KYS)~\cite{jap}. Using
diagrammatic and effective potential methods in a particular
MSSM, as well as various
approximations, Hempfling and Hoang (HH)~\cite{hemp}
found that two-loop corrections are also
sizeable, but negative with respect to the one-loop result!
Using the RG approach,
we have found~\cite{CEQR}
that two-loop corrections are negative with respect to the
one-loop result. We have traced back the origin of this
disagreement with KYS~\cite{jap} in their choice of the minimization scale.
Furthermore KYS neglected various effects
(as the contribution of gauge bosons to the one-loop effective potential,
or the wave function renormalization of top quark and Higgs boson)
and considered only the case with zero stop mixing.
On the other hand, we have found that the
abnormal size of the two-loop corrections obtained
by HH~\cite{hemp}
is a consequence of an excessively rough estimate of the one-loop
result, but we are in agreement with their final two-loop result.
In fact our two-loop results differ from those of HH~\cite{hemp}
by less than $3\%$. Also our results show a large
sensitivity of the Higgs mass to the stop mixing parameter.

Finally we would like to comment briefly on the generality of our
results. As was already stated, we are assuming average
squark masses $M_S^2\gg M_Z^2$,
and that all supersymmetric particle masses are $\simgt M_S$. If
we relax the last assumption,
i.e. if some supersymmetric particles were much
lighter, the value of the quartic coupling at $M_S$ (see Eq.~(\ref{lms}))
would be slightly increased and, correspondingly,
our bounds would be slightly relaxed. We have made an
estimate of this effect. Assuming an extreme case where all gauginos,
higgsinos and sleptons have masses $\sim M_Z$,
we have found for $M_S=1$ TeV and
$\cos^2 2\beta=1$ an increase of the Higgs mass
$\sim 2\%$. For values of $\tan\beta$ close to 1 (as those appearing in
infrared fixed point scenarios) the corresponding effect is negligible.
On the other hand, our numerical results
have been computed for $M_S = 1$ TeV. For values of
$M_S < 1$ TeV the bounds on the lightest Higgs mass
are lowered. Hence, in this sense, all our
results can be considered as absolute upper bounds.

\section{Conclusions}

We want to conclude with comments on the two questions we raised at the
beginning of this talk: {\bf i)} Can the Higgs mass measurement disentangle
between the SM and the MSSM? {\bf ii)} Can
eventually a Higgs mass measurement at
LEP-200 help in putting upper bounds on the scale of new physics?

As for question {\bf i)}, a quick glance at Fig.~\ref{comparacion}
shows that for $M_t=173\pm 4$ GeV, i.e.
the crossing area of the SM
and MSSM curves, the Higgs mass eventually measured will be
compatible either with the pure SM  or with the MSSM, but not with
both at the same time.
Accordingly, the experimental Higgs mass either will discard the MSSM
or will be a clear signal of new physics beyond the SM compatible
with the MSSM.
For $M_t<169$  GeV, the situation is analogous, but there is a
wider range of Higgs masses (area within the two curves) compatible
with both SM and MSSM.
For $M_t>177$ GeV, there is no region of
Higgs masses simultaneously compatible
with the SM and MSSM. On the contrary there is a range
of $M_H$ (within the two curves) that would discard both. This means
that for $M_t>M_t^c=177$ GeV a Higgs mass measurement could {\bf always}
discriminate between the SM and the MSSM. This statement has two caveats:
one is that it is based upon assuming that the SM holds up to the Planck
scale. Had we considered the validity of the SM up to a lower scale, we
should have obtained a value $M_t^c>177$ GeV, with a larger overlapping
region where the Higgs mass measurement cannot discriminate between the
SM and MSSM. The second caveat is that we are requiring absolute stability
for the effective potential: instead, the requirement of metastability of
the electroweak minimum will decrease the lower bound and go along the same
direction as the previous effect. A detailed calculation along these lines
is at present being performed.

\begin{figure}[htb]
\centerline{
\psfig{figure=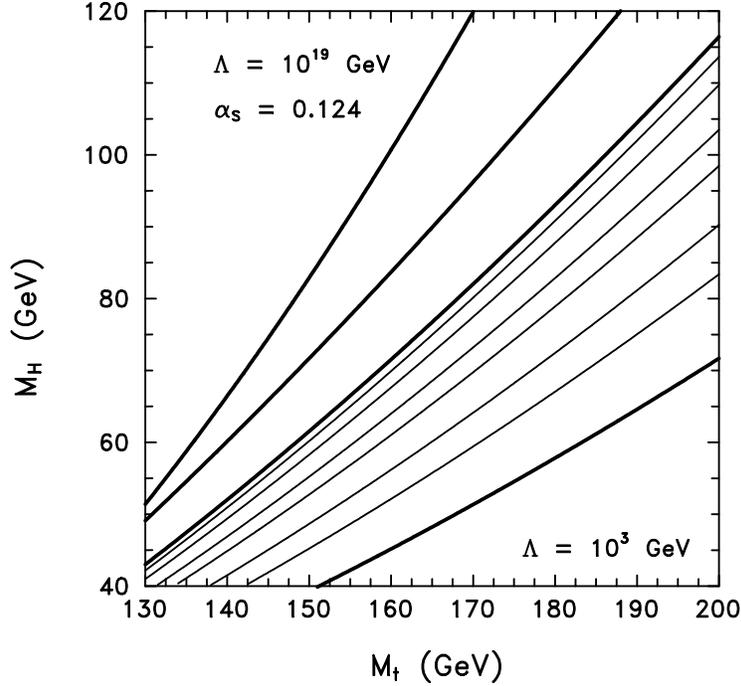,height=10cm,bbllx=5cm,bblly=5cm,bburx=18cm,bbury=17cm}}
\caption[0]{SM lower bound on $M_H$ as a function of $M_t$ for
$\alpha_s(M_Z)=0.124$ and different values of $\Lambda$ in the range
$10^3$ GeV $\leq \Lambda \leq 10^{19}$ GeV. Thick lines correspond
to $\Lambda=10^3,\ 10^4,\ 10^5$ and $10^{19}$ GeV. Thin lines
correspond to $\Lambda\equiv$1.5, 2, 3, 4, 6 and 8~TeV.}
\label{detalle}
\end{figure}

As for question {\bf ii)}, we can see from Fig.~\ref{detalle} that an
upper bound on the scale of new physics can be deduced from the Higgs
mass measurement. For instance, if $M_H<M_Z$, and fixing $M_t\sim 175$
GeV, we obtain from
Fig.~\ref{detalle} that $\Lambda \simlt 10$ TeV.
This bound crucially depends on the value of the top mass and will be
softened by the second (metastability bound) effect
just mentioned.

\section{Acknowledgements}
I want to express my deepest gratitude to my collaborators on the
topics covered by this talk, Alberto
Casas, Jos\'e Ram\'on Espinosa and Antonio Riotto.
I also acknowledge discussions with
Marcela Carena, Gordy Kane,
Santi Peris, Nir Polonsky, Stephan Pokorski, Carlos
Wagner and Fabio Zwirner. I finally wish to thank Fabio Zwirner for a
careful reading of the manuscript.

\end{document}